\documentclass[10pt,conference]{IEEEtran} 

\usepackage[ruled]{algorithm2e}
\usepackage{times}                               
\usepackage{graphicx}
\usepackage{tabularx}
\usepackage{mathtools}
\usepackage{algo}
\usepackage{amsmath}
\usepackage{amssymb}
\usepackage{units}
\usepackage{textcomp,xspace}
\usepackage{enumerate}
\usepackage{multirow}
\usepackage{multicol}
\usepackage{subfigure}
\usepackage{color}
\usepackage{xcolor}
\usepackage{url}
\usepackage{authblk}
\definecolor{lightgray}{gray}{0.9}

\DeclarePairedDelimiter{\ceil}{\lceil}{\rceil}

\newcommand{\subw}{\ensuremath{\text{sub}}}
\newcommand{\compl}{\ensuremath{\text{comp}}}
\renewcommand{\algorithmcfname}{ALGORITHM}

\begin{document}

\title{Optimal Codeword Construction for DNA-based Finite Automata}

\author[$\dagger$]{Anupam Chattopadhyay}
\author[$\ddagger$]{Arnab Chakrabarti}
\affil[$\dagger$]{School of Computer Science and Engineering, Nanyang Technological University, Singapore}
\affil[$\ddagger$]{Uniklinik Aachen, RWTH Aachen University, Germany}

\maketitle

\begin{abstract}
Biomolecular computation has emerged as an important area of computer science research due to its high information density, immense parallelism opportunity along with potential applications in cryptography, genetic engineering and bioinformatics. Computational frameworks using DNA molecules have been proposed in the literature to accomplish varied tasks such as simulating logical operations, performing matrix multiplication, and encoding instances of NP-hard problems. In one of the key applications, several studies have proposed construction of finite automata using DNA hybridisation and ligation~\cite{soreni_jacs}\cite{benenson_first_fsm}. The state and symbol encoding of these finite automata are done manually. In this manuscript, we study the codeword construction problem for this approach. We derive exact theoretical bounds on the number of symbols and states in the finite automata and also obtain the complete set of symbols in a specific case, thereby solving an open problem posed in~\cite{soreni_jacs}. For automatic encoding, two different solutions, based on a heuristic and on Integer Linear Programming (ILP), are proposed. Furthermore, we propose an early simulation-based validation of laboratory experiments. Our proposed flow accepts a finite automaton, automatically encodes the symbols for the actual experiments and executes the simulation step-by-step.
\end{abstract}

\sloppy

\section{Introduction}
The field of DNA computing was initiated by the landmark work of Adleman~\cite{adleman} who showed how to solve small instances of the NP-complete {\em Hamiltonian Path} problem by encoding them on DNA strands. DNA, or Deoxyribonucleic Acid, is a fundamental unit that encodes the genetic information that makes life possible. It is a polymer whose monomeric units consist of four nucleotides: Adenine (A), Cytosine (C), Guanine (G) and Thymine (T). The letters of this alphabet bind according to the well-known Watson-Crick complement condition: Adenine binds to Thymine (A $\leftrightarrow$ T) and Cytosine binds to Guanine (C $\leftrightarrow$ G).

\begin{figure}[htb]
\begin{center}
    \includegraphics[angle=0,width=60mm]{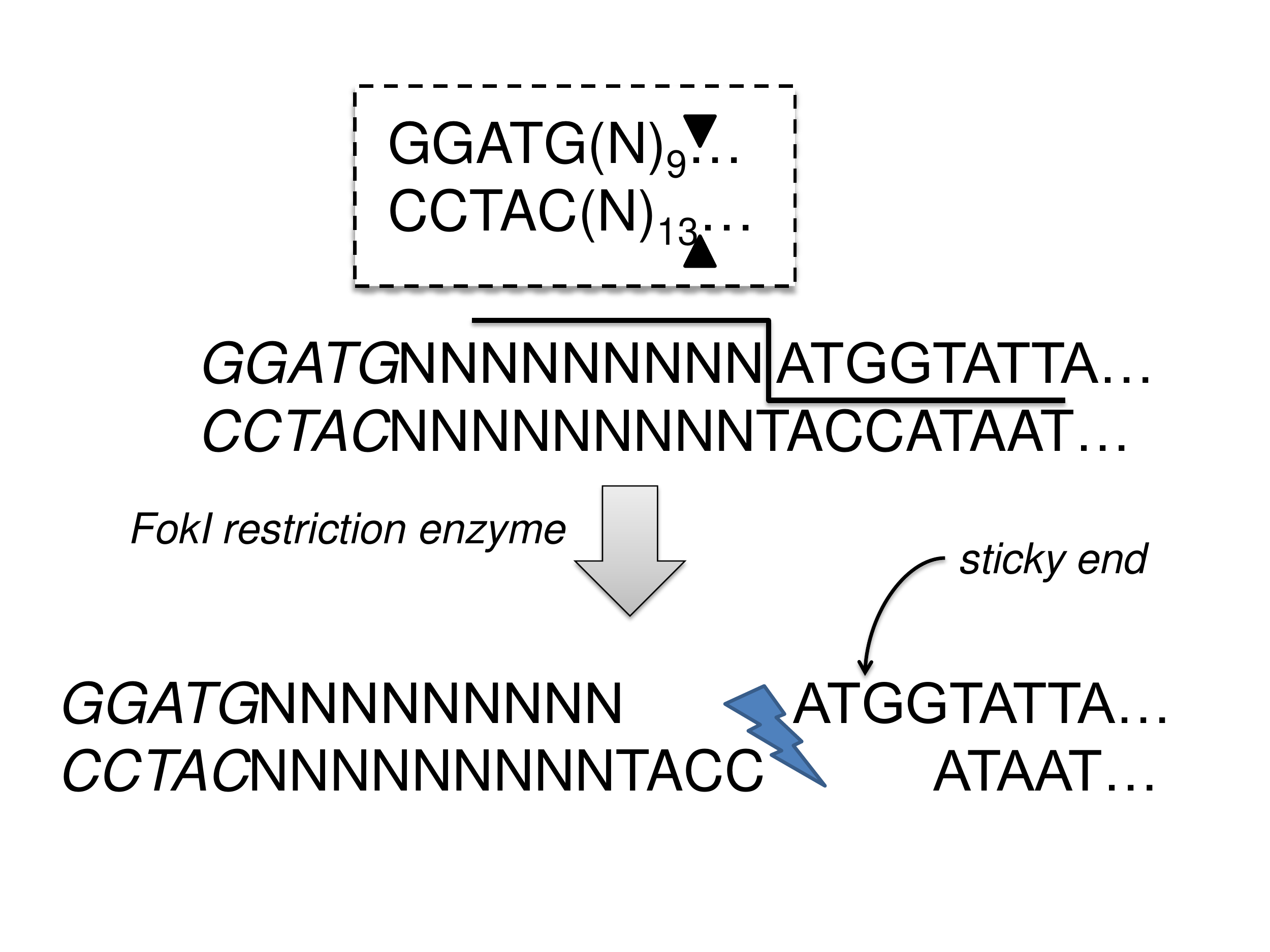}
\vspace{-0.8cm}
    \caption{Restriction Enzyme Operation}
    \label{fig:restr-enzyme}
\end{center}
\vspace{-0.8cm}
\end{figure}

Typically, to use DNA molecules to perform computations, problem instances have to be encoded onto a set of DNA molecules. These molecules are then allowed to react with or without the influence of suitable enzymes to form potential solutions. Finally one uses protocols such as gel electrophoresis to obtain solutions in molecular form. Some of these biomolecular reactions include:

\textit{Hybridisation:} A double-stranded DNA molecule is formed by creating hydrogen bonds between the complementary bases of two single-stranded molecules that have opposite orientations.

\textit{Polymerase Enzyme:} Polymerase are DNA enzymes which help duplicate a DNA strand. A Polymerase Chain Reaction (PCR) is an oft-used protocol for detecting or amplifying a specific molecule within a large mix of molecules.

\textit{Ligation:} If a double stranded DNA molecule contains a broken bond in one of the strands then it can be glued by a \textit{ligase} enzyme such as $T4$. The process is known as ligation.

\textit{Restriction Endonuclease/Enzyme:} A restriction enzyme recognizes a particular sequence of nucleotides in a double stranded DNA and cuts it into two pieces by destroying the phosphodiester bonds at a specific distance from the recognition site. This is essentially the inverse operation of ligation. The single strands hanging out of each piece are known as \emph{sticky ends}. The recognition site as well as the distance from it to the cutting point are inherent properties of a restriction enzyme~\cite{neblab,szy_univ}. An example of such as a reaction via the restriction enzyme $FokI$ is shown in Figure \ref{fig:restr-enzyme}. $FokI$ cuts the base strand from its recognition site at a distance of nine nucleotides and the complementary strand at a distance of thirteen nucleotides producing a sticky end consisting of four nucleotides. Several restriction enzymes can also perform blunt cuts which do not produce any sticky end.

\textit{Gel Electrophoresis:} This technique allows separating the DNA molecules according to their weight (or length). This is used for controlling the intermediate steps of a reaction or for reading the final output.

In the original paper by Adleman~\cite{adleman}, a number of the aforementioned steps were employed. The high information density and massive parallelism of DNA computing sparked major research interest in this field, though it was shown that DNA computing does not offer any additional power from the perspective of computability theory. Several NP-hard problems were subsequently encoded into DNA molecules~\cite{lipton_np_complete} including the well-known {\em Circuit Satisfiability} problem~\cite{boneh_dna_power,liu_3sat}. Several important practical applications followed including the breaking of the Data Encryption Standard (DES)~\cite{boneh_des} and the possibility of parallel algorithms for integer factorization~\cite{chang_factoring}, a key step in several public key cryptography protocols. This initiated the field of DNA-based cryptography~\cite{gehani_dna_crypto}. A detailed survey of DNA computing is presented in~\cite{amos_dna_book}. A serious disadvantage of the encoding approaches proposed in~\cite{adleman,boneh_dna_power} is that the number of DNA strands required to encode instances grows exponentially with their size (see Table~1 in~\cite{boneh_dna_power}). There have been some advances in encoding problem instances of modest sizes. For instance, Braich et al.~\cite{braich_20sat} show how to solve a twenty-variable {\em $3$-SAT} instance. However encoding large problem instances so that all potential solutions are simultaneously constructed, as is the approach in~\cite{braich_20sat}, is difficult since, molecular computations are inherently error-prone and this affects the reliability of the final result~\cite{boneh_error_res}. Recent research in DNA computing has, therefore, focused on the construction of small nanodevices that can be scaled up for solving practical problems~\cite{Qian2014,Roquetaad8559}.

\subsection{Finite Automata using DNA Computing}
The computing power of biomolecular nanodevices that are capable of self-assembly was studied by Winfree et al.~\cite{winfree_universal}. They showed that there can be different self-assembly primitives for DNA molecules that can express regular grammars (finite automata), context-free grammars and unrestricted grammars. In particular, it was shown that a $2$-dimensional self-assembly of DNA molecules can simulate a Universal Turing Machine. This study spear-headed experimental work to validate these models.

Benenson et al.~\cite{benenson_first_fsm} were the first to propose constructing a finite state automaton with DNA molecules by using restriction enzymes. Their model was simple enough to be used in a genetic control study~\cite{benenson_gene_control} with far-reaching implications in genetic engineering, pharmaceutical experiments and medical diagnosis. Subsequently, experiments with transducers i.e. finite automata with outputs, have been reported in~\cite{banani_transducer} and there has been some success in $2$-dimensional self-assembly of DNA tiles~\cite{winfree_sierpinsky}. Based on experimentally validated finite automata constructions, novel theoretical models for push-down automata have also been proposed~\cite{jonoska_automata,krasinski_pda,cavaliere_unbounded_2005}.

The simplicity of the finite automata construction using restriction enzymes motivated researchers to scale up the problem size. Experimental studies with an autonomous, programmable $3$-symbol, $3$-state finite automata was reported in~\cite{soreni_jacs}. The authors of this paper claimed that their approach can be used to construct a $37$-symbol, $3$-state finite automata (a bound that we improve in this paper). This design approach led to interesting applications in molecular image encryption by using a $2$-state, $2$-symbol state machine~\cite{shosani_image}.

\subsection{Codeword Design for DNA Computing}\label{ssec:dna-code-related}
Codeword construction for DNA is an well-studied topic due to its importance in tasks like DNA-based information storage, DNA computing or molecular bar code reading for chemical libraries. A good codeword construction aim towards increasing information density and/or parallelism in computing without creating cross-hybridisation between distinct codewords. The goal of maximizing the information density can be reduced to the problem of finding large sets of DNA strands maximally separated in a metric space, where strands with high hybridization affinity are mapped to neighboring points. In this setting, it is proved that codeword design is an NP-complete problem~\cite{Phan_NP_complete}. Consequently, various heuristics have been proposed for DNA codeword design, including generic algorithm~\cite{genetic_search}, taboo search~\cite{taboo_search}, stochastic search~\cite{stochastic}, and also drew inspiration from the traditional problem of binary codeword construction~\cite{elgamal}. A survey of DNA codeword construction challenges and approaches is presented in~\cite{Codeword_survey}.

Often DNA codeword design entails satisfying much stronger constraints than enforced by pure non-hybridization~\cite{hartemink_scan}\cite{demo_word_design}. Such combinatorial constraints often originate from the use case, e.g., DNA-based self-assembly or finite automata construction, as we focus in this manuscript. Depending on the experimental approach, different combinatorial constraints and different codeword design strategy are needed, as described in, e.g.,~\cite{tadeusz_arithmetic}\cite{zhang_constraint}.

In this work, we focus on the codeword construction for finite automata using a single restriction enzyme, as proposed in~\cite{soreni_jacs}. To the best of our knowledge, this has not been attempted before, thus leaving several interesting research problems unresolved. The problem of codeword construction for a similar setting, with multiple restriction enzymes, is discussed in~\cite{tadeusz_arithmetic}. For recombinase-based state machines (RSMs), the codeword construction problem is recently proposed in~\cite{RSM_science}.

It is also imperative to prepare a simulation environment before undertaking the actual experiment. A generic simulation flow for DNA computing was developed in~\cite{ms_dna_simulator} where one can model and simulate various bio-molecular reactions. However, the problem of constructing distinguished languages (the encoding problem) was not addressed there. A design automation environment for Boolean circuit construction via DNA computing was developed in~\cite{qian_seesaw}.

In this perspective, our main contributions in this paper are outlined below.
\begin{itemize}
\item For arbitrary $p$ and $l$, we identify theoretical bounds on the number of valid words and the size of a $(p,l)$-distinguished language. This allows us to determine the maximum number of symbols and $(\text{state},\text{symbol})$ pairs possible for restriction enzymes-based constructions of finite automata.
\item We design efficient algorithms that enumerate an optimal or a near-optimal sized $(p,l)$-distinguished language.
\item We also develop a simulation flow to allow \emph{dry} experiments and validation.
\end{itemize}

The rest of the paper is organized as follows. Section~\ref{sec:preliminaries} introduces the formal structure of the problem. Section~\ref{sec:fa_example} contains a detailed example of a finite automaton construction using DNA molecules. The encoding problem is dealt with in Section~\ref{sec:fa_encoding} and we report experimental results about the efficiency of our automation and the construction of finite automata with a larger state count than reported in~\cite{soreni_jacs} in Section~\ref{sec:exp}. Section~\ref{conclusion} contains some concluding remarks. Supplementary material with the actual symbol enumeration is presented in Appendix.

\section{Preliminaries and Motivation}\label{sec:preliminaries}
In this section, we briefly state some of the basic definitions and terminology that are used in this paper.

A \emph{finite state machine (FSM)} is a quintuple $(\Sigma, S, s_0, \delta, F)$, where
\begin{itemize}
\item $\Sigma$ is the input alphabet (a finite, non-empty set of symbols);

\item $S$ is a finite, non-empty set of states;

\item $s_0 \in S$ is the initial state;

\item $\delta \colon S \times \Sigma \rightarrow S$ is the state-transition function;

\item $F \subseteq S$ is the set of final states (possibly empty).
\end{itemize}

To simulate an FSM using DNA molecules, we encode its symbols using the alphabet $\Sigma = \{A, C, G, T\}$ which comprises of the letters representing the nucleotides
that constitute DNA. Note that there are \emph{two} alphabets under consideration here: the alphabet of the FSM and the alphabet that we use to encode the states and symbols of the FSM. To avoid confusion, we use the term \emph{symbol} to denote an element of the input alphabet of an FSM and \emph{letter} to denote an element of $\Sigma$, the encoding alphabet.

In what follows, fix $\Sigma = \{A,C,G,T\}$. Given a positive integer~$p$, we denote by $\Sigma^p$ the set of all words of length~$p$ from the alphabet~$\Sigma$. The set of all words from $\Sigma$ is denoted by $\Sigma^{*}$. For $w = a_1a_2 \cdots a_p \in \Sigma^p$ and $i \leq j$, define $w[i,j]$ to be the subword of $w$ that starts at index~$i$ and ends at index~$j$, that is, $w[i,j] = a_i a_{i+1} \ldots a_j$. Given an integer $1 \leq l \leq p$, we let $\subw_l(w)$ denote the set of $p-l+1$ subwords $w[1,l], w[2,l+1], \ldots, w[p-l+1,p]$ of length $l$. Given a word $w = a_1 \cdots a_r \in \Sigma^{*}$, we define its \emph{complement} to be the word $\bar{w} = \bar{a}_1 \cdots \bar{a}_r$, where $\bar{a}_i$ is the Watson-Crick complement of the letter $a_i$, that is, A and T are complements and so are C and G. To capture this notion in general, we introduce the notion of a \emph{paired alphabet}. A \emph{pairing} $P \subseteq \Sigma \times \Sigma$ of an alphabet $\Sigma$ is defined as a set of tuples such that every $a \in \Sigma$ appears in exactly one tuple once. We then call $(\Sigma,P)$ a \emph{paired} alphabet---in our settings this will be given by $(\{A,C,T,G\},\{(A,T),(C,G),(T,A),(G,C)\})$. For sake of readability and a slight abuse of notation, we will use $\Sigma$ to denote a paired alphabet in the following. Under this definition, the notion of complements is easily extensible by choosing the letter-wise complement as the pairs appearing in the pairing and using the above extension to words. We then denote by $\compl_l(w)$ the set of complements of the words in the set $\subw_l(w)$. A word $w = a_1,\ldots, a_p \in \Sigma^{p}$ is a \emph{palindrome} if for all $1 \leq i \leq \lfloor p/2 \rfloor$, $a_i$ is the Watson-Crick complement of $a_{p - i +1}$.

Note that this definition of a palindrome is different from the usual definition for words where the letters $a_i$ and $a_{p - i +1}$ are required to be identical. In DNA computing, one is interested in preventing nucleotide strands to fold and form hairpin loop (alternatively known as stem-loop) structures~\cite{watson_book}. In the domain of microbiology research such palindromic sequences are also referred to as Inverted Repeat (IR)~\cite{ussery_book}.

Let us fix integers $p$ and $l$ with $1 \leq l \leq p$. A word $w \in \Sigma^p$ is \emph{$(p,l)$-valid} if it satisfies the following conditions:
\begin{enumerate}
\item $\left | \subw_l(w) \cup \compl_l(w) \right | = 2 \cdot (p - l + 1)$;
\item no word in $\subw_l(w)$ is a palindrome.
\end{enumerate}
The first condition says that the length-$l$ subwords of $w$ are distinct and no two of them are complements of one another. Thus a word is valid if its subwords are distinct, are not palindromes and no two of them are complements of one another. Formally, this can be captured with the notion of \emph{$(p,l)$-distinguished} language. We call a language $L \subseteq \Sigma^{p}$, \emph{$(p,l)$-distinguished} if
\begin{enumerate}
\item every word of $L$ is $(p,l)$-valid, and,
\item for all $w,w' \in L$ with $w \neq w'$ we have that $$(\subw_l(w) \cup \compl_l(w)) \cap (\subw_l(w') \cup \compl_l(w')) = \emptyset.$$
\end{enumerate}

In this definition, the second condition states that distinct words in a distinguished language neither have the same subwords nor their complements. Intuitively speaking, a \emph{valid word} in our setting is used to encode a \emph{single} input symbol of the finite automaton under construction. Subwords are used to encode the $(\emph{state},\emph{symbol})$ pairs of the automaton. As will be clear later on, the restrictions on words and languages that were described above allow us to encode inputs to the automaton in such a way that one can use restriction enzymes to simulate its computation on the given input. While this approach was outlined by Soreni et al.~\cite{soreni_jacs}, they manually determined words that encode input symbols.\footnote{Further, the symbol table reported in~\cite{soreni_jacs}, Table 1 contains overlapping 4-bp sequences between No. 11 and No. 18. Thus all codewords are not valid.} This leads to the open problem that, what is the maximum achievable number of valid words for a \emph{$(p,l)$-distinguished} language and how to determine those words. For a better understanding of the problem scope, we now look closely into an exemplary FSM.

\begin{figure*}[htb]
\begin{center}
    \includegraphics[angle=0,width=120mm]{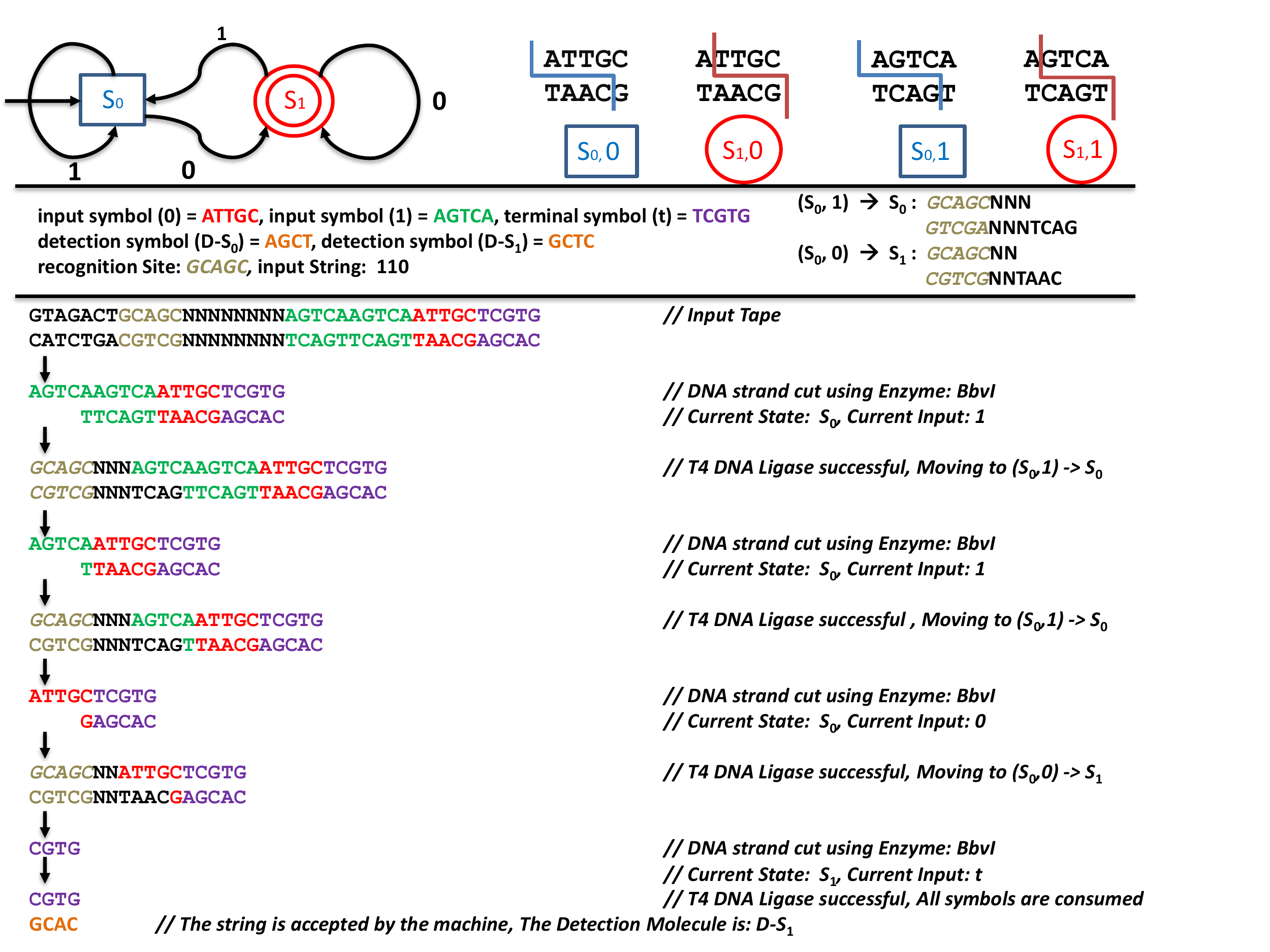}
    \caption{FSM via DNA Operations}
    \label{fig:start-fsm}
\end{center}
\end{figure*}
\vspace{-0.2cm}

\section{An Example FSM} \label{sec:fa_example}
Consider the finite automaton shown in Figure~\ref{fig:start-fsm} which accepts binary strings ending with $0$. In this section we revisit the simulation such an automaton using DNA molecules~\cite{benenson_first_fsm}. As was explained in Section~\ref{sec:preliminaries}, we use the alphabet $\Sigma = \{A, C, G, T\}$ to encode input symbols of an automaton.

In this example we set $p = 5$ and $l = 4$ and encode the input symbols $0,1$ by valid $(5,4)$-words $w_0 = \text{ATTGC}$ and $w_1 = \text{AGTCA}$, respectively. We also encode a terminating symbol $t$ by the valid $(5,4)$-word TCGTG making sure that $\{w_0,w_1,t\}$ is a $(5,4)$-distinguished language. For $i \in \{0,1\}$, the subword $w_i[1,4]$ represents the fact that the automaton is in state~$S_0$ and reading symbol~$i$; and, the subword $w_i[2,5]$ denotes that the automaton is in state~$S_1$ and reading symbol~$i$. In this way, all possible $(\text{state},\text{symbol})$ pairs are encoded. Since $\{w_0,w_1,t\}$ is $(5,4)$-distinguished, the subwords $w_0[1,4], w_0[2,5], w_1[1,4], w_1[2,5],t[1,4], t[2,5]$ are all distinct, are not palindromes and no two of these are complements of one another.

The input to the automaton is encoded on the \emph{base strand} of a double-stranded DNA molecule (shown on top in Figure~\ref{fig:start-fsm}). In this instance, the string $110t$ is encoded in the base strand along with additional information to guide the action of the restriction enzyme which in this case is $BbvI$. The second strand, called the \emph{complementary strand}, consists of complementary nucleotides that align with their partners in the base strand. Together these two strands constitute our input tape. The restriction enzyme $BbvI$ has the property that it recognizes a sequence of nucleotides (the recognition site) which in this case is the sequence GCAGC. It cuts the base strand eight nucleotides after the recognition site and the complementary strand twelve nucleotides after the recognition site. When this happens for the first time, it exposes the sequence AGTC on the base strand which as discussed above means that the automaton is in state $S_0$ reading the symbol~$1$. This exposed sequence is called the \emph{sticky end} because it is in the nature of nucleotides to form bonds with their complementary partners. At this point, the T4 ligase attaches itself to this exposed end via the complementary sticky end TCAG as shown in the figure. This operation, in effect, deletes the first input symbol from the input tape because when the restriction enzyme works on it again, it cuts the base strand beyond the first input symbol ignoring it in the process. This sequence of cutting the strands by the restriction enzyme followed by ligation continues until the whole of the input is ``consumed.'' When the terminal symbols are exposed, special nucleotide sequences called \emph{detection symbols} attach themselves to them. Depending on which sequence of terminal symbols were exposed by the restriction enzyme, one can detect whether the automaton terminated in state~$S_0$ or~$S_1$. In an actual experiment, one determines the termination state by additional biochemical reactions such as polymerase chain reactions and gel electrophoresis. In the example shown, the automaton terminates in state~$S_1$ as it should.

\section{Automated State Encoding} \label{sec:fa_encoding}
Having explained how one might construct an FSM using DNA strands and restriction enzymes, we find it convenient to explain some the requirement on the words and languages from $\Sigma^{*}$ that we imposed in Section~\ref{sec:preliminaries} in order to be able to encode inputs to the FSM. In the context of $(p,l)$-distinguished languages, the subwords $\subw_l(w)$ are the sticky ends produced by the restriction enzymes and these represent the $(\text{state},\text{symbol})$ pairs of the automaton.

\begin{description}
\item[\emph{rule 1:}] The sticky ends, produced by the restriction enzymes, cannot be palindromes as they would stick to one another to form \emph{hairpin loop}~\cite{watson_book,bikard_hairpin}.

\item[\emph{rule 2:}] Each sticky end represents a particular $(\text{symbol},\text{state})$ pair. To uniquely identify a symbol, a sticky end cannot occur more than once within the symbol itself or within any other symbols.

\item[\emph{rule 3:}] The complement of the sticky end, selected for a state of a symbol, cannot occur in the same or any other symbol.

\item[\emph{rule 4:}] All the DNA symbols should have a sequence which is unique. So there cannot be a repetition of similar DNA symbols in the library of valid symbols.
\end{description}

Note that rule 4 is encompassing rule 2. The mapping of valid symbols to the states are followed by fixed choices for the symbols of state transitions. On the basis of these constraints, it is possible to identify the theoretical bounds on the symbol and state count, which is done in the following sub-section. A specific case of this bound calculation is presented in~\cite{soreni_jacs}.

\subsection{An Upper bound for FSM encoding} \label{sec:Bound}
\def\lhalf{{\lfloor l/2 \rfloor}}

Assume that $L_{p,l}$ is a $(p,l)$-distinguished language of maximal cardinality. A simple upper bound for $|L_{p,l}|$ can be obtained as follows: the total number of non-palindrome $l$-letter subwords is given by $|\Sigma|^l - |\Sigma|^\lhalf$. Excluding complement subwords, at most $\frac{1}{2}(|\Sigma|^l - |\Sigma|^\lhalf)$ such $l$-letter words can appear as subwords in $L_{p,l}$.

As every $p$-letter word in a $(p,l)$-distinguished language contains exactly $p-l+1$ distinct subwords of length $l$, we obtain the bound
$$
	|L_{p,l}| \leq \frac{|\Sigma|^l - |\Sigma|^\lhalf}{2 (p - l + 1)}.
$$

Surprisingly, this very simple limit is tight for the range of parameters that we explored so far (cf. Table~\ref{table:ilp_heu}) and we suspect that it is
tight for all $p,l$, except for the case of $l$ being $1$.

For a given FSM $(\Gamma, S, s_0, \delta, F)$ and parameter $p,l$ we then obtain the following bounds:
\begin{align}
	|\Gamma| &\leq |L_{p,l}| - 1 \\
	|S| &\leq p-l+1  			
\end{align}
which in turn bounds the number of transitions from above by $(|L_{p,l}|-1)(l-p+1)$. Note that we cannot use every word of $L_{p,l}$ to encode symbol-state pairs, as
the construction reserves one terminal-symbol to read out the result of the computation.

\subsection{Problem Formulation}
Let us first formalize the optimization problem of finding a maximal $(p,l)$-distinguished language over a paired alphabet $\Sigma$.

\begin{tabular*}{\textwidth}{@{\hspace{.5em}} >{\itshape}l p{0.3\textwidth} @{}}%
                \multicolumn{2}{l}{\hspace{0.0em}\textsc{Max $\Sigma$-Distinguished Language}} \vspace{.2em}\\
	Input: & Integers $p,l$ \\	
	Problem: & Compute a $(p,l)$-distinguished language of maximal cardinality\\
\end{tabular*}

We suspect that this problem can be solved in time polynomial in $|\Sigma|^p$, but this remains an interesting open question. For the scope of this application we provide a heuristic that solves this problem with reasonable solution quality.

\begin{figure}[hbt]
\begin{center}
    \includegraphics[angle=0,width=90mm]{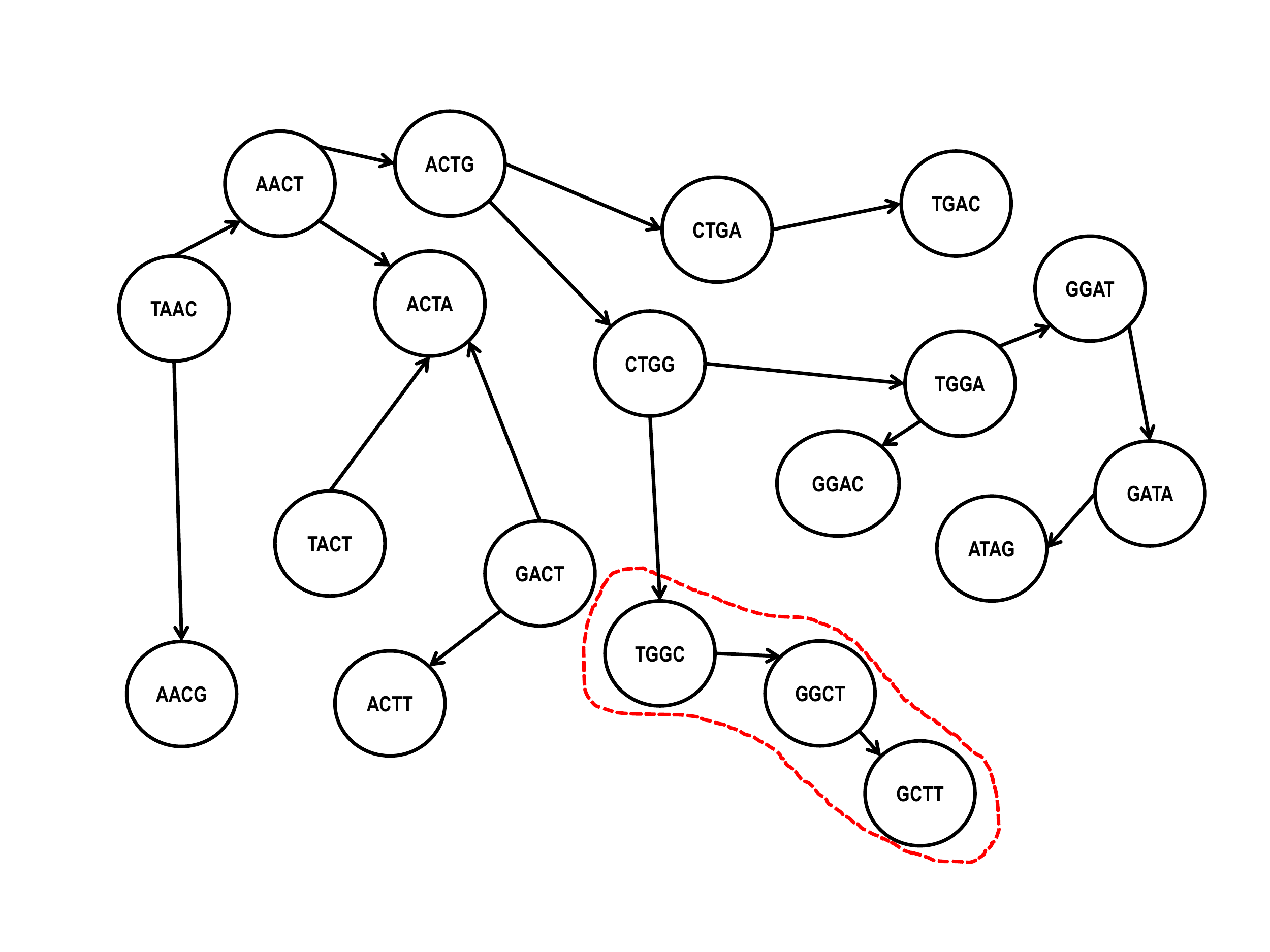}
    \caption{FSM Encoding Problem}
    \label{fig:p3-cover}
\end{center}
\end{figure}

The aforementioned constraints for valid symbols can be conveniently represented in a graphical view, as shown in the Figure \ref{fig:p3-cover}. In the graph,
the $4$-letter symbols made out of the alphabets $\{A, T, G, C\}$ are nodes ($V$). A node, $V_1$ is connected via a directed edge ($E$) with $V_2$ only if the last $3$
letters of $V_1$ are exactly same as the first $3$ letters of $V_2$. The graph constructed satisfies following properties.
\begin{itemize}
 \item The graph contains several cycles. Exact enumeration of cycle count is also possible but, not required for this work.
 \item The maximum indegree of any vertex ($V_i$) is $\leq$ $4$.
 \item The maximum outdegree of any vertex ($V_i$) is $\leq$ $4$.
\end{itemize}

\vspace{-0.2cm}

From Figure \ref{fig:p3-cover}, the enumeration of all valid symbols can be alternatively viewed as enumeration of all directed, vertex-disjoint node covers of length $3$, also known as $P3$-cover. In the general case, the number of nodes to be covered is dependent on the value of $S_{max}$ identified in subsection \ref{sec:Bound}. For the number of states $S_{max}$ being equal to $\textit{n}$, the problem is to determine maximum, vertex-disjoint $P\textit{n}$-covers.


\subsection{Heuristic Approach} \label{algo:heu}
For the heuristic computation, we transform an instance $(\Sigma,p,l)$ into a directed auxiliary graph $G_A$ as follows: the vertices of $G_A$ are exactly the non-palindrome $l$-letter words of $\Sigma^l$ and two words $w,w' \in V(G_A)$ are connected by an arc $(w,w')$ if $w[2,l] = w'[1,l-1]$. Thus, a directed path in $G$ of length $p-l+1$ corresponds to a word in $\Sigma^p$ without repeated or palindrome subwords---note that such a word might be invalid, as it could contain both a subword and its complement.

The heuristic now greedily chooses directed $(p-l+1)$-paths such that all vertices on that path have a low out-degree. Every vertex can be used at most once in this process, thus no $p$-letter subword will appear twice. However, as a word and its complement could be chosen, we need a post-processing step to create a valid $(p,l)$-distinguished language.

\begin{algorithm}[htbp]
\KwIn{$p$, $l$}
\KwOut{$L$}
\caption{Heuristic via Directed Path-cover}
$L = \emptyset$\;
$G_A = createGraph(p,l)$\;
$degree = 0$\;
\While{$G_A$ contains vertices with outdegree $\ge degree$} {
$v = $ Vertex of minimal outdegree $\ge$ degree\;
$P = (v)$\;
\For{$i = 1 \to (p-l) $} {
$w = $ Out-neighbour of $v$ with minimal outdegree\;
$P = P + w$\;
$v = w$\;
}
\eIf{$length(P) = p$}{
$L = L \cup word(G_A,P)$\;
$G_A = G_A \setminus P$\;
}{
$degree = degree + 1$\;
}
}
$G_c = createConflictGraph(L)$\;
\While{$G_c$ contains node with degree $\ge 1$}{
$w = $ Vertex of maximum degree in $G_c$\;
$G_c = G_c - w$\;
$L = L \setminus \{w\}$\;
}
\end{algorithm}

At first, the algorithm proceeds by creating an exhaustive non-palindromic list of symbols with length $l$. The graph is then constructed by creating directed edges when there is an overlap of literals between $2$ nodes. The first while-loop combines neighbouring symbols as long as possible. This is controlled by an increasing outdegree. At the beginning, all nodes are checked and a symbol is formed by connecting a node with minimal outdegree to its neighbouring node, which also has minimal outdegree and so on. The nodes are connected till the valid symbol size ($p$) is reached, which is alternatively indicated by the maximum state count ($S_{max}$ = $p$ - $l$ + $1$). In case, the formation of a valid symbol is not possible starting from the indicated node, the outdegree is increased by one and the iterative search for a valid symbol begins again. If no such node is found, the outdegree requirement is increased by $1$ and the While-loop is iterated. This guarantees that the algorithm is terminated when no more symbol construction is possible. After every valid symbol generation, the corresponding nodes are removed from the graph.

The last part of the algorithm enforces the rule 3 mentioned in the section~\ref{sec:fa_encoding}. During symbol construction, complementary sticky ends may become part of the same or different symbols. This is identified by an exhaustive pairwise check of all the symbols and then a conflict graph is constructed. In the conflict graph, a conflict edge denotes that either of the connected nodes (i.e. symbols) can exist in the valid symbol library. To obtain maximum number of symbols, repeatedly nodes with maximum degree are removed till no conflict edge remains.

\subsection{Algorithmic Complexity}
The worst-case execution time for the above heuristic is based on the assumption that all combinations of $S_{max}$ nodes are tried for constructing a valid symbol. In practice, this is less since, the choice of the first node bounds the algorithm to search for a following node within at most $4$ options. This makes the total number of possible combinations as $n \cdot 4^{S_{max}-1}$, where $n$ is the number of nodes in the graph $G_A$. For $n$ being $(4^{l} - 4^{\ceil{l/2}})$, the complexity is $\mathcal O(4^{l} \cdot 4^{S_{max}-1})$. A similar complexity figure of $\mathcal O(4^{l} \cdot 4^{l})$ is obtained for the second part of the algorithm, where conflict edges are removed. Despite the exponential complexity, the heuristic turned out to be highly efficient for the currently used restriction enzymes where, $l$ $\le$ $4$ as well as several other scenarios presented in the section~\ref{sec:exp}.

\subsection{ILP Based Approach} \label{algo:ilp}
Whereas the above heuristic yields a large set of codewords in a reasonable amount of time, a provable \emph{maximal} set of codewords could be desirable. To this end we reduce the problem of finding a maximal set of codewords with word length $p$ and subword-length $l$ to finding a independent set in a suitable \emph{conflict graph}. As this problem is NP-hard in general, this approach can only work for a limited range of parameters. However, using this formulation we were able to calculate--- among other parameters---an optimal number of codewords for $p=6$ and $l=4$.

Let us describe the construction of the \emph{conflict graph} $G_{p,l}$ in detail: the vertices of $G_{p,l}$ are exactly the $(p,l)$-valid words of $\Sigma^p$ and
we add an edge between $w,w' \in V(G_{p,l})$ if
$$
	(\subw_l(w) \cup \compl_l(w)) \cap (\subw_l(w') \cup \compl_l(w')) \neq \emptyset
$$
holds. That is: two words $w,w'$ are connected by an edge in the conflict graph if not both can be in a $(p,l)$-distinguished language $L \subseteq \Sigma^p$ at the same time. Conversely, any set $X \subseteq V(G_{p,l})$ that is independent, i.e. $G_{p,l}[X]$ contains no edges, is by definition $(p,l)$-distinguished. Thus a maximal $(p,l)$-distinguished language can be found by finding a maximum independent set in $G_{p,l}$.

As \textsc{Maximum Independent Set} is a well-established problem, we leave away the ILP-formulation that can easily be derived from a given conflict graph $G_{p,l}$. A maximum set of $(6,4)$-distinguished words of cardinality $40$ that was obtained solving \textsc{Maximum Independent Set} on the conflict graph $G_{6,4}$ can be found in the Appendix.

\subsection{Transition Encoding}
We explain the transition encoding by taking an exemplary transition from $state_m$ to $state_n$ with the input symbol being $sym$. The double-stranded transition molecule has the format of $\{$\textit{recognition}$\_$\textit{site} $\|$ \textit{buffer}$\_$\textit{strand} $\|$ \textit{complementary}$\_$\textit{single}$\_$\textit{strand}$\}$. The \textit{complementary}$\_$\textit{single}$\_$\textit{strand} is determined from the encoding of the $state_m$, which remains in the sticky end after every cut by restriction enzyme. The length of the \textit{buffer}$\_$\textit{strand} is allocated in such a way that the next cut by restriction enzyme consumes the current symbol $sym$ and reaches the corresponding $state_n$ of the next symbol. For every symbol, the state encodings are done in a regular fashion. Therefore, the $state_n$ of next symbol always requires consumption of a fixed amount of literals. Note, that the encoding for \textit{recognition}$\_$\textit{site} and the \textit{buffer}$\_$\textit{strand} must be different from the encoding of the symbols.

\begin{figure}[hbt]
\begin{center}
    \includegraphics[angle=0,width=90mm]{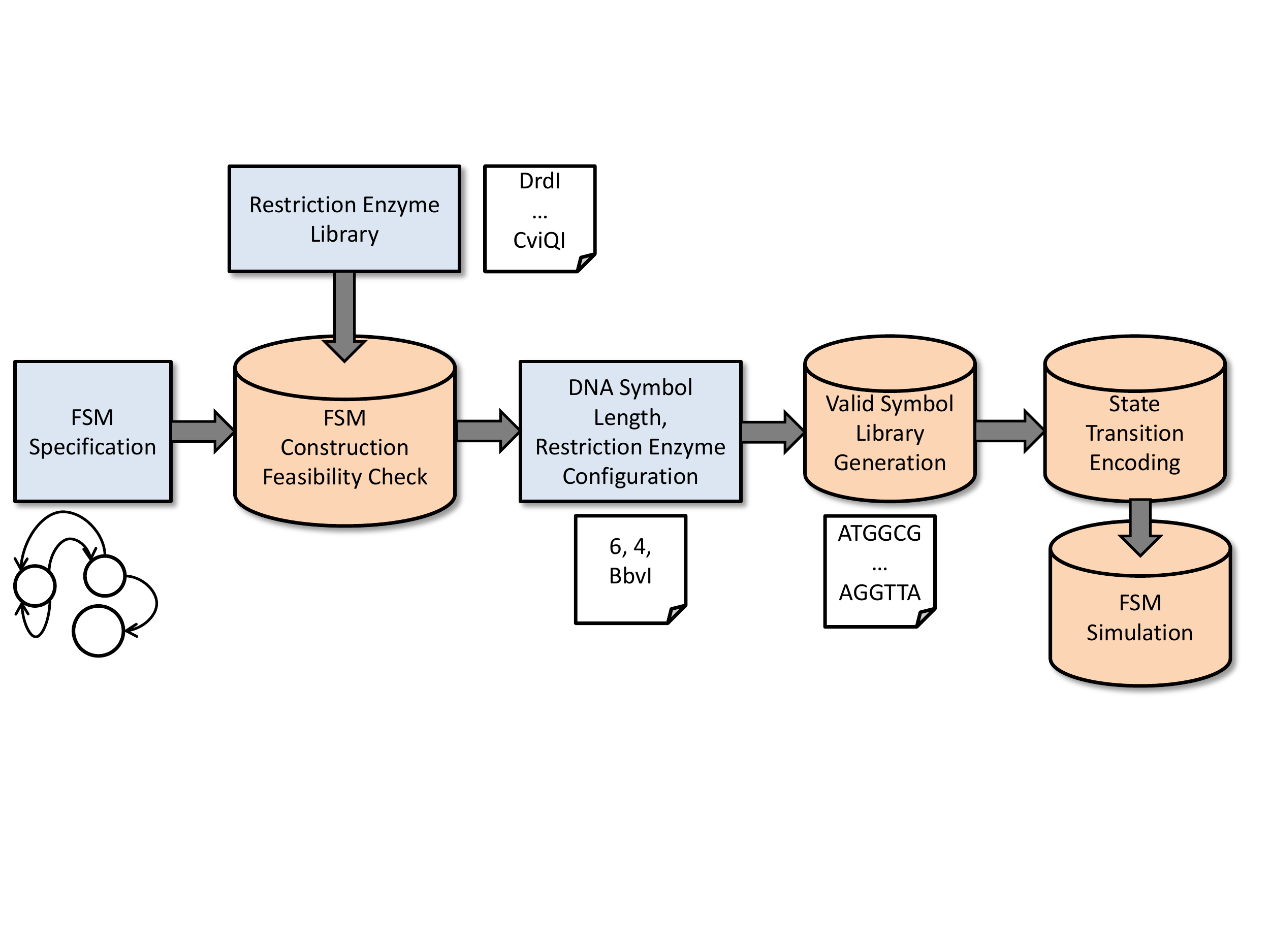}
\vspace{-2.0cm}
    \caption{Tool Flow for FSM Encoding and Simulation}
    \label{fig:sim-flow}
\end{center}
\end{figure}
\vspace{-0.2cm}

\subsection{Software Work flow} \label{simulation_flow}
The algorithms presented in the previous section is packaged in a software tool flow.  The software follows the steps outlined in Figure~\ref{fig:sim-flow}. It accepts as input a set of restriction enzymes~\cite{neblab} and an FSM. First, it checks if it is feasible to construct the DNA-automaton. If so, an appropriate encoding of the FSM is automatically generated using the symbols taken from the restriction enzymes provided. Finally, the finite automaton is simulated step by step for a given input tape.



\section{Experimental Evaluation} \label{sec:exp}
The complete tool flow reported in this paper is developed using C++, consisting of about 2600 lines of code. To solve the ILPs the CPLEX solver was used \cite{cplex}.  The code was tested on a AMD Phenom II X6 1100T 3.3 GHz 6-core processor with 8GB RAM running 64-bit scientific Linux OS. We will present two sets of experiments: In the first one, a simulation of a 5-state FSM using our tool is shown as an example and the second is a benchmark measuring the efficiency of generating optimal or near-optimal symbols for state encoding.

\subsection{Finite Automata for Checking Divisibility by 5}
\begin{figure}[hbt]
\begin{center}
\includegraphics[angle=0,width=90mm]{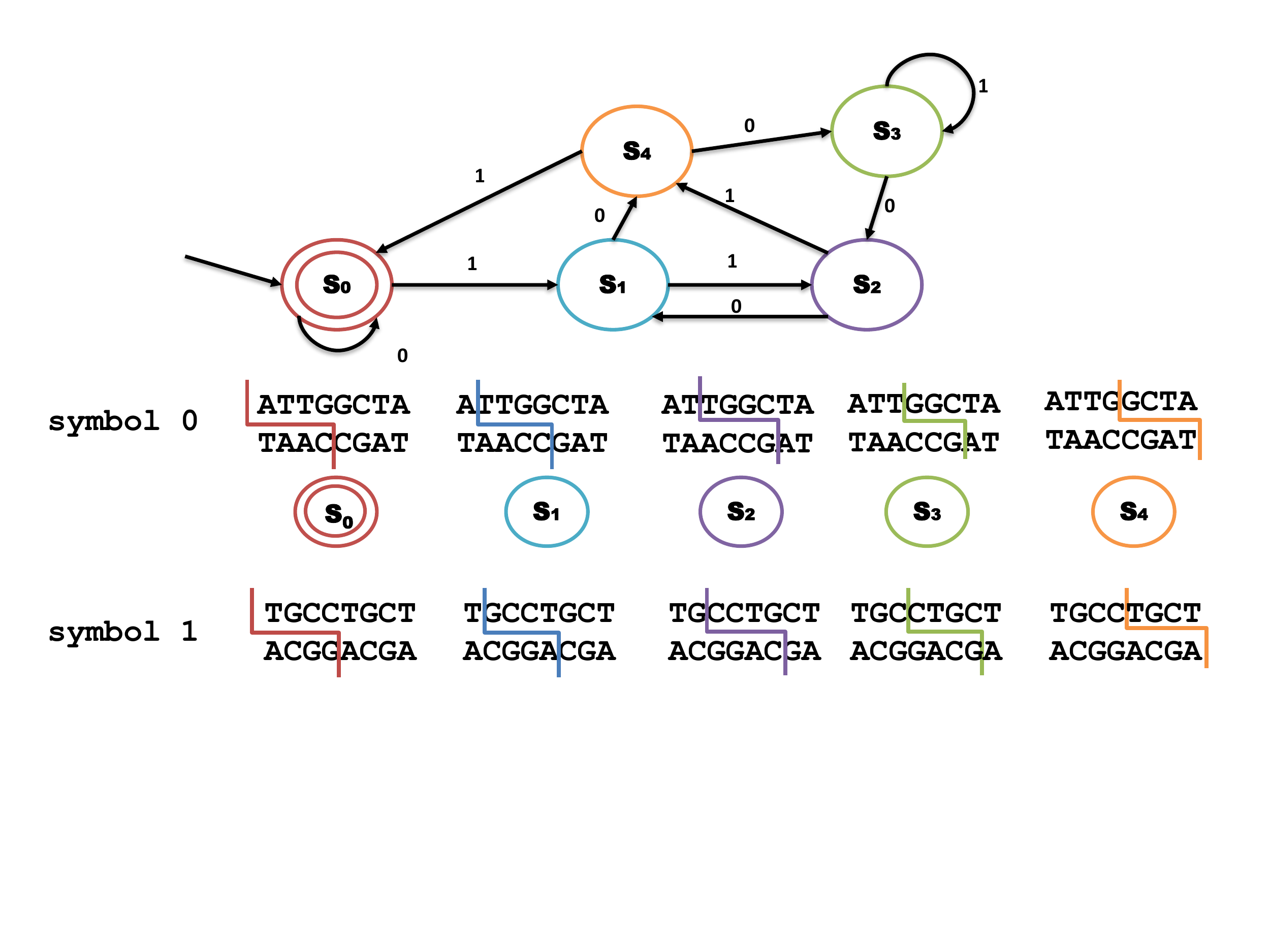}
\vspace{-2.0cm}
   \caption{FSM for Divisibility by 5}
   \label{fig:div5}
\end{center}
\vspace{-0.7cm}
\end{figure}

With the capabilities provided by our tool, that allows efficiently, automated encoding, it becomes possible to execute FSMs with a higher number of states and symbols than reported earlier. For example, one can construct a $3$-state, $39$-symbol FSM using the restriction enzyme $BbvI$ (we actually have $40$ symbols available, but we need to use one to signalize termination).  For a realistic non-trivial example an FSM construction for divisibility check is chosen. The automata accepts binary strings of $0$ and $1$ that represent numbers that are multiples of five. The automata, that can be seen in Figure~\ref{fig:div5}, has a total of $5$ states and $2$ symbols. This means we have to use a symbol length of $8$ for the restriction enzymes with a sticky end of length $4$. For creating the transition encoding, which can consume the complete symbol, one needs a restriction enzyme having a distance from recognition site to the cutting point to be $\ge$ $9$. If this is not the case, then we might not be able to cut enough to transition between different states. This is automatically determined by the tool and a matching restriction enzyme $FokI$ is selected. The state
encoding and the transition encoding are automatically determined within $1$ second. The detailed simulation steps for different binary strings can be found in Appendix. It should also be noted that FSM is a powerful construction that can be applied to many different problems, two more cases of which are also presented in the Appendix.

\subsection{Benchmarking Symbol Generation Efficiency}
To determine the efficiency of our proposed algorithms for generating the state symbols, different values of $p$ and $l$ are fed to the algorithm. The corresponding theoretical maximum are computed. The generated ILP and heuristic based symbol counts for different $p$ and $l$ values are presented in Table~\ref{table:ilp_heu}. The ILP-based method could not scale for certain values due to the increasing number of ILP constraints. But as can be seen from the theoretical upper bound this is not important, since the number of possible symbols decreases with growing $p$ for a fixed $l$. This may seem counter-intuitive at first. The reason is that, with such a small alphabet, it becomes increasingly difficult to not repeat a subword with increasing word length. Furthermore, the right enzymes must exist to be able to use certain $(p,l)$ pairs. To the best of our knowledge, none that would work with larger DNA strings than the ones for which we present results are known. For the parameter values with acceptable ILP constraints, the pre-computed maximum number of symbols is always returned. In contrast, the heuristic approach scales well and returns symbols, which are quite close to the optimal. The heuristic approach took less than $1$ second to generate the symbols in all the presented cases. For ILP, the timeout was set to $1$ hour, within which several scenarios failed to return the symbols.

\begin{table}[htbp]
\centering
\caption{Enumeration of State and Symbol Counts}
{\begin{tabular}{cccccc}
\hline
\multirow{2}{*}{$p$}    &  \multirow{2}{*}{$l$} & \multirow{2}{*}{State Count} & \multicolumn{3}{c}{Symbol Count} \\ \cline{4-6}
 & & & Upper bound & ILP & Heuristic \\ \hline
6 & 4 & 3 & 40 & 40 & 38 \\
7 & 4 & 4 & 30 & 30 & 26 \\
7 & 5 & 3 & 160 & 160 & 142 \\
8 & 4 & 3 & 24 & \textemdash & 20 \\
8 & 6 & 5 & 672 & 672 & 558 \\
9 & 4 & 6 & 20 & \textemdash & 18 \\
10 & 4 & 7 & 17 & \textemdash & 12 \\
\hline
\label{table:ilp_heu}
\end{tabular}}
\end{table}
\vspace{-0.4cm}

\section{Conclusion and Future Work} \label{conclusion}
In this paper, an automated and efficient finite state machine encoding approach is presented. The optimal solutions for the number of achievable symbol and state were generated to be used in the our tool. To find such solutions, an efficient heuristic and an ILP-based approach are proposed, which are verified through simulation.

\section*{Acknowledgement}
Authors will like to gratefully acknowledge the help of Felix Reidl, Fernando S\'{a}nchez Villaamil and Somnath Sikdar, who were affiliated with Theoretical Computer Science, RWTH Aachen University at the time, and helped with the ILP formulation.

\bibliographystyle{IEEEtran}
\bibliography{dna}

\newpage

\appendix
\section*{Optimal Set of Symbols for $6$-letter symbol with Enzyme BbvI} \label{optimal_sym}

\begin{table}[ht]
\centering
\begin{tabular}{|c c c c c|}

\hline
CGAAAA & ATGGAA & CCTGAA & TGACAA & TCGTAA \\
ACCAGA & CGAGGA & GACGGA & GAACGA & ATCCGA \\
CTATGA & ATGTGA & TAAACA & TAGCCA & CGCCCA \\
GATTCA & CAGGTA & GGTGTA & TAACTA & TCCCTA \\
CCTCTA & GTATTA & TAGTTA & CGGCAG & ACTCAG \\
AAATAG & CAGTAG & TCTTAG & GCAAGG & AGTGGG \\
GGTTGG & CGCACG & ACAGCG & GGGGCG & AATGCG \\
CGACCG & ATCTCG & GTTTCG & GCGATG & CTTCTG \\

\hline
\end{tabular}
\label{table:sym40}
\end{table}


\begin{figure}[hbt]
\begin{center}
   \includegraphics[angle=0,width=90mm]{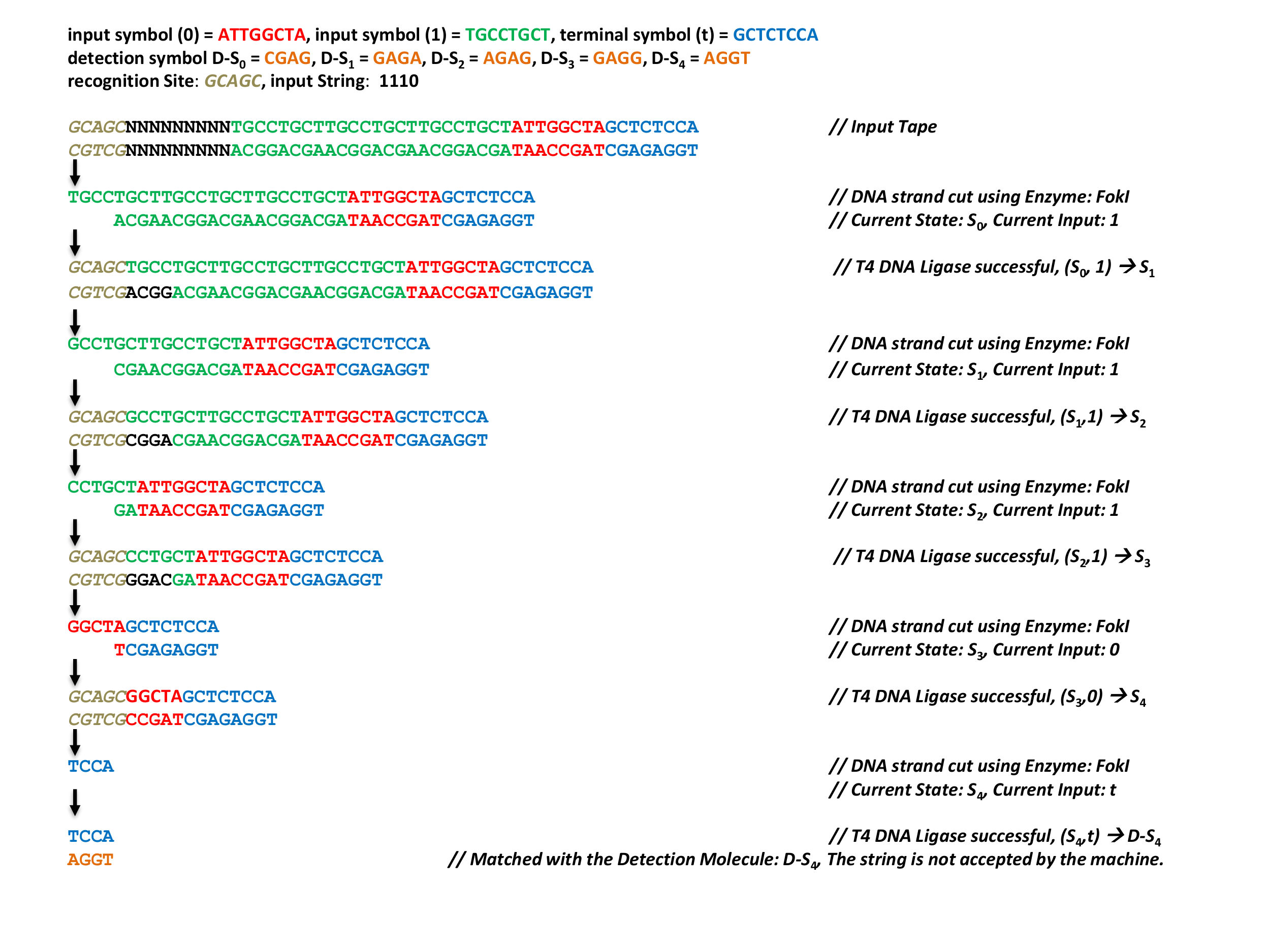}
   \caption{Divisibility by 5: FSM Simulation Steps for Binary Word 1110}
   \label{fig:div5_noa}
\end{center}
\end{figure}




\begin{figure}[hbt]
\begin{center}
   \includegraphics[angle=0,width=90mm]{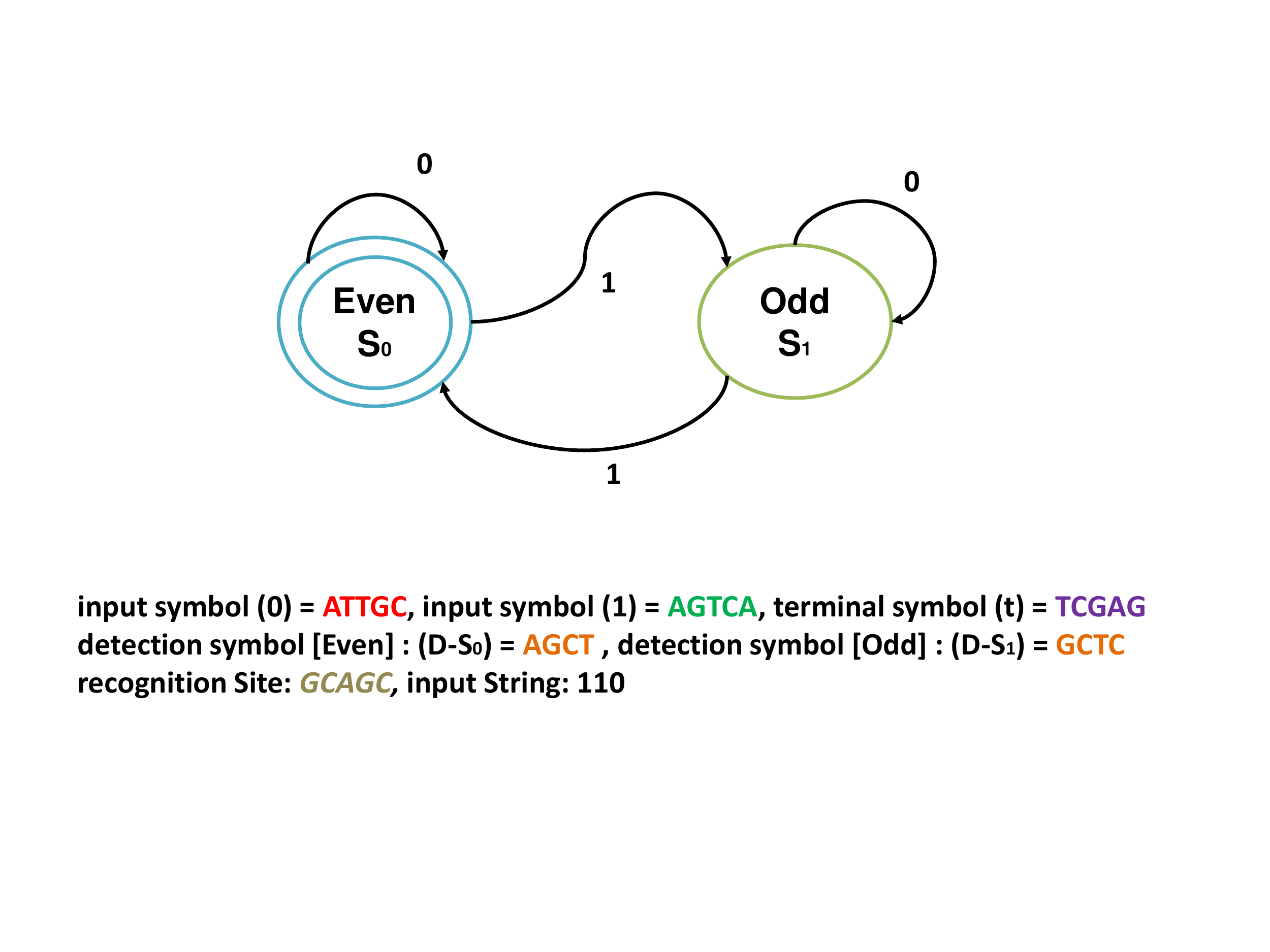}
   \caption{FSM for Parity Checker}
   \label{fig:parity-fsm}
\end{center}
\end{figure}

\begin{figure}[hbt]
\begin{center}
   \includegraphics[angle=0,width=90mm]{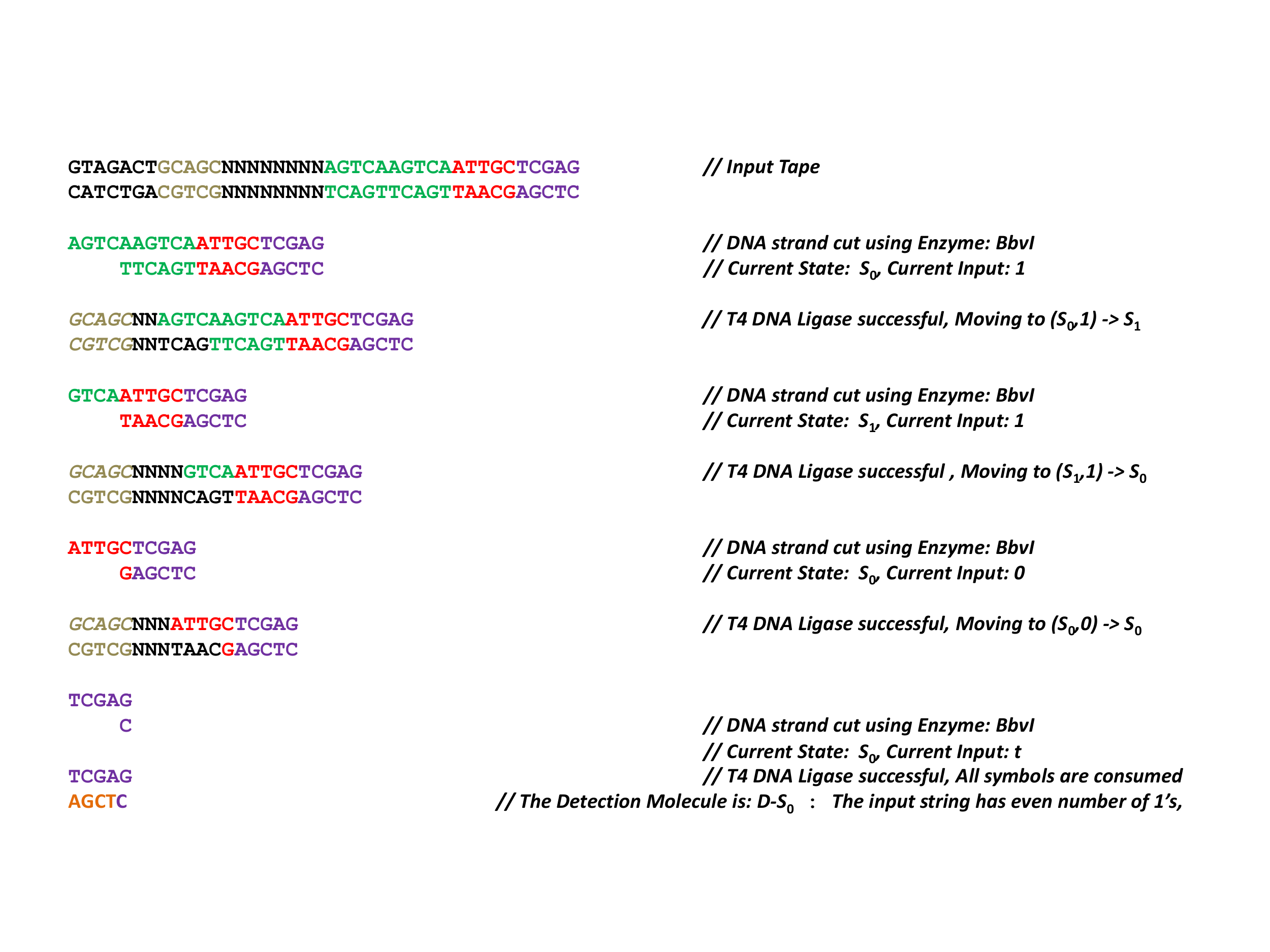}
   \caption{FSM Simulation Steps for Parity Checker}
    \label{fig:parity-steps}
\end{center}
\end{figure}


\begin{figure}[hbt]
\begin{center}
   \includegraphics[angle=0,width=90mm]{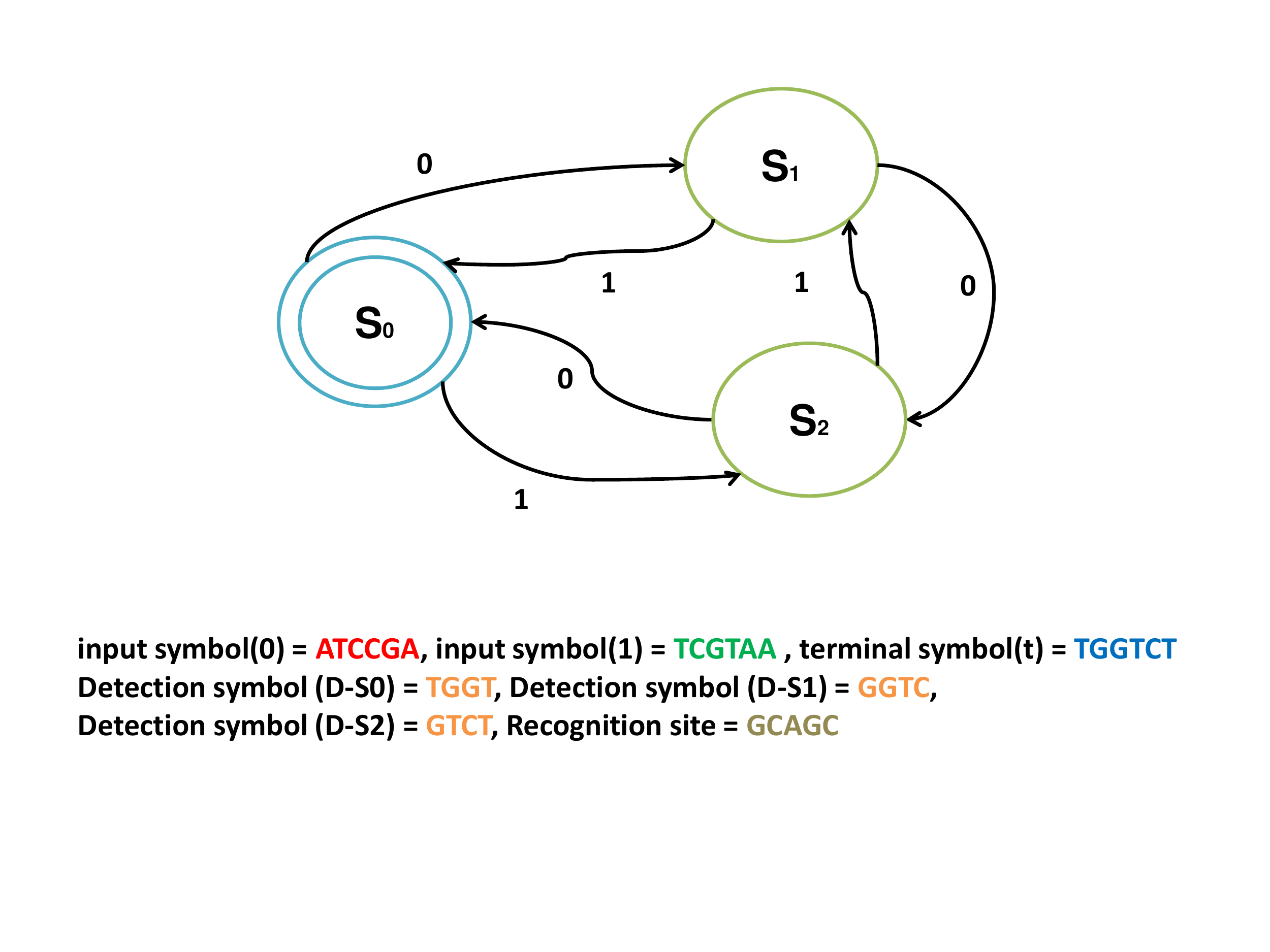}
   \caption{FSM for Regular Expression Matching}
   \label{fig:regex-fsm}
\end{center}
\end{figure}

\begin{figure}[hbt]
\begin{center}
   \includegraphics[angle=0,width=90mm]{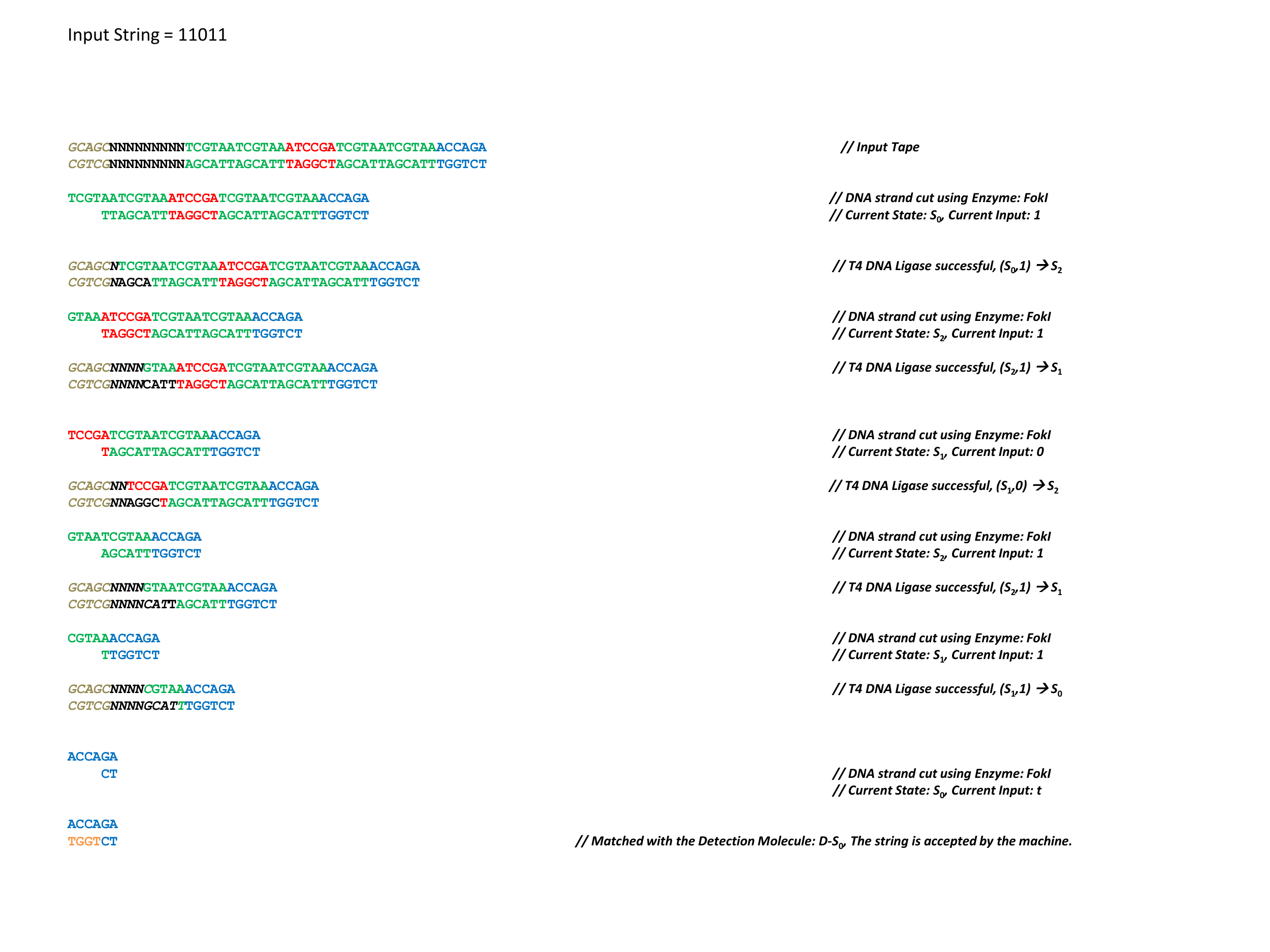}
   \caption{FSM Simulation Steps for Regular Expression Matching}
    \label{fig:regex-steps}
\end{center}
\end{figure}

\end{document}